\newcommand{\Msun}{~M_\odot}

\newcommand{\lsun}{L_\odot}

\newcommand{\kms}{\rm ~km~s^{-1}}
\newcommand{\ergs}{\rm ~erg~s^{-1}}

\newcommand{\ml}{~\Msun ~\rm yr^{-1}}

\documentclass[preprint]{aastex}

\begin{document}

\title{THE NATURE OF THE COMPACT SUPERNOVA REMNANTS IN STARBURST GALAXIES}
\author{Roger A. Chevalier}
\affil{Department of Astronomy, University of Virginia, P.O. Box 3818, \\
Charlottesville, VA 22903; rac5x@virginia.edu}
\author{Claes Fransson}
\affil{Stockholm Observatory, Department of Astronomy,
SCFAB, \\
 SE-106 91 Stockholm, Sweden;
claes@astro.su.se}


\begin{abstract}
Radio observations of starburst regions in galaxies have revealed
groups of compact nonthermal sources that 
may be radiative supernova remnants expanding in the
interclump medium of molecular clouds.
Because of the high pressure in starburst regions, the interclump medium
may have a density $\sim 10^3$ H atoms cm$^{-3}$ in a starburst nucleus
like M82 and $\ga 10^4$ H atoms cm$^{-3}$ in an ultraluminous galaxy
like Arp 220.
In M82, our model can account for the sizes, the slow evolution,  the
high radio luminosities, and the low X-ray luminosities of the sources.
We predict expansion velocities $\sim 500\kms$, which is slower
than the one case measured by VLBI techniques.
Although we predict the remnants to be radiative, the expected radiation is
difficult to detect because it is at infrared wavelengths and the starburst
is itself very luminous; one detection possibility is  broad [OI] 63 $\mu$m
line emission at the positions of the radio remnants.
The more luminous and compact remnants in Arp 220 can be accounted for
by the higher molecular cloud density.
In our model, the observed
 remnants lose most of the supernova energy to radiation.
Other explosions in a lower density medium may
directly heat a hot, low density interstellar component, leading to
the superwinds that are associated with starburst regions.

\end{abstract}


\section{INTRODUCTION}

Radio observations of starburst galaxies have revealed a population
of compact supernova remnants that are more luminous 
than Cas A, the most luminous supernova remnant in our Galaxy.
This population has been best studied in M82 (Kronberg, Biermann, \&
Schwab 1985), and similar populations have been found in
NGC 253 (Antonucci \& Ulvestad 1988),
NGC 3448 (Noreau \& Kronberg 1987),
and  Arp 220 (Smith et al. 1998).
Although there is general agreement that these nonthermal radio
sources are the result of supernovae, there has been no clear sense of
their evolutionary status.
One possibility is that they involve interaction with the winds from
the progenitor massive star.
The radio supernova phenomenon can be interpreted in terms of supernova
interaction with the free wind of the progenitor (Chevalier 1982;
Weiler et al. 1986).
In this case, the radio properties do not depend on the interstellar
environment.
Alternatively, the remnants are interacting with interstellar material,
in which case the starburst environment is likely to be important.

Recent observations of the compact remnants in M82 have yielded clues
about their nature.
Kronberg et al. (2000) found that 18 of the sources (75\%
of the sample) showed no
discernible flux evolution over a 12 year period, which suggests a
flux variability $<0.1$\% yr$^{-1}$ (see also Ulvestad \& Antonucci
1994).
The evolution timescale is then $>10^3 |n|$ yr if the luminosity
evolves as $L\propto t^n$.
An estimate of $n$ can be obtained from the radio $\Sigma - D$
(surface brightness -- diameter) relation, assuming that it
represents an evolutionary trend; for $\Sigma \propto D^{-3.5}$
(Huang et al. 1994), we have $|n|=1.5$.
Muxlow et al. (1994) resolved most the remnants in M82, finding
diameters of $0.6-4$ pc, and suggested that they are supernova remnants
rather than radio supernovae.
In some cases, shell structure was observable, as expected for
a supernova remnant.
The combination of size and timescale yields a velocity $\la 10^3\kms$,
which implies that these are evolved remnants.
This forms the basis of our treatment of the compact remnants
in M82 (\S~2) and in other galaxies (\S~3).
We also briefly discuss supernova-driven superwinds in \S~3.
We take a distance to M82 of 3.63 Mpc (Freedman et al. 1994).

\section{THE COMPACT REMNANTS IN M82}

Of the 24 radio sources in M82 followed by Kronberg et al. (2000),
two, 41.5+59.7 and 41.9+57.5, show especially rapid evolution.
The first one was observed in 1981, but not in 1982.
The rapid evolution was consistent with the
radio emission from a Type Ib or Ic supernova (Kronberg et al.
2000), and we agree that
this is a plausible interpretation.
The emission is from a shock wave propagating in the free wind
from the progenitor star.
The other source, which has been the brightest of the compact sources,
had an evolution that is compatible with that of a bright radio supernova like
SN 1986J, but recent radio imaging shows that it has a bipolar structure
(McDonald et al. 2001).
We do not include 41.9+57.5 in our discussion.

The duration of the initial radio supernova phase is limited by the
extent of the freely expanding progenitor wind.
In the case of a starburst region, the wind expansion is likely to be
limited by the high pressure of the interstellar medium.
Based on energy input from supernovae, Chevalier \& Clegg (1985)
estimated a pressure in the central region of M82: 
$p/k\approx 10^7$ cm$^{-3}$ K, where $k$ is Boltzmann's
constant.
The ram pressure of a wind equals the external pressure at a
radius
$r_w=0.2\dot M_{-4}^{1/2}v_{w1}^{1/2}p_7^{-1/2}$ pc,
where $\dot M_{-4}$ is the mass loss rate in units of $10^{-4}\ml$,
$v_{w1}$ is the wind velocity in units of $10\kms$,
and $p_7$ is the external pressure in units of $10^7$ cm$^{-3}$ K.
Except for 41.9+57.5, the radio remnants have
$r\ga 0.5$ pc (Muxlow et al. 1994), which is too large for interaction
with a red supergiant wind.
A Wolf-Rayet progenitor wind could expand to a larger radius because of the
larger wind velocity ($v_{w1}\approx 100$), but the density is low,
leading to a faint source.
The majority of
the compact sources in M82 cannot be interpreted as
standard radio supernovae.

The interstellar medium in the center of M82 is believed to have a hot
component that drives a galactic superwind (Chevalier \& Clegg 1985;
McCarthy, van Breugel, \& Heckman 1987; Strickland \& Stevens 2000).
In the reference wind model of Chevalier \& Clegg (1985) with
a power of one $10^{51}$ erg supernova every $3-5$ years and
a mass loss rate of $1\ml$, the central temperature is
$T_c\approx 5\times 10^8$ K and central H density is
$n_H\approx 0.05$ cm$^{-3}$.   
The models of Strickland \& Stevens (2000) are cooler, and appear
to have $n_H\sim 0.1$ cm$^{-3}$ in the central region.
In {\it Chandra} observations,
Griffiths et al. (2000) found an energetic X-ray component
in the central kpc of M82 that is more luminous than predicted
by the thermal emission in the wind models, but much of the
emission could be from the inverse Compton mechanism. 
The radius at which a supernova has swept up its own mass is
$R_s=8.9(M/10\Msun)^{1/3}(n_H/0.1{\rm~cm^{-3}})^{-1/3}$ pc,
where $M$ is the ejecta mass, and the radius at which a $10^{51}$ erg
supernova comes into pressure equilibrium with its surroundings
is $R_p=16 p_7^{-1/3}$ pc.
Under these conditions, the observed radio remnants, with diameters
$\la 4$ pc (Muxlow et al. 1994),
would be primarily in free expansion, with
characteristic velocities $\ga 5,000\kms$.

In view of the large evolutionary times for the compact remnants,
we propose that most of the compact remnants are interacting with
a dense, possibly molecular,  interstellar medium.
The analogy in our Galaxy would be remnants evolving in the
interclump medium of molecular clouds (Chevalier 1999), which
has a H density $n_H\approx 10$ cm$^{-3}$.
Galactic remnants like IC 443 and W44, which have high radio surface
brightnesses and radii $\sim 10$ pc, can be interpreted 
as recently having entered the
radiative phase of evolution.
The remnants in M82 can also be interpreted as recently having
entered the radiative phase if the surrounding density is
$n_H\approx 10^3$ cm$^{-3}$.
This is a plausible density for the interclump medium of molecular
clouds in M82 if the high pressure in the central region is
taken into account.
In our Galaxy, the pressure of the interclump medium of clouds is
maintained at $\sim 10^5$ cm$^{-3}$ K by nonthermal support,
presumably related to magnetic fields.
In M82, a similar situation is likely.
Shen \& Lo (1995) find a CO line width of $7\kms$ and up in
molecular clouds in M82.
When this line width is combined with a density $n_H\approx 10^3$ cm$^{-3}$,
the result is a nonthermal pressure $p/k\sim 10^7$ cm$^{-3}$ K, as is
inferred in the central region of M82.
In addition, there is direct evidence for densities $\ga 10^3$ H atoms
cm$^{-3}$ in the interclump medium of molecular clouds in
starburst regions (Mao et al. 2000; Solomon 2001).
The location of the radio remnants in M82 overlaps the regions of
strong CO emission (e.g., Fig. 4 of Lord et al. 1996) and the
gas column density to the remnants is high (Mattila \& Meikle 2001),
which is consistent with interaction with molecular gas.

A possible complication 
is that the analysis of the dense gas emission in terms of
photodissociation regions implies that the gas has a filling
factor $\sim 0.05-0.1$ and is in many
clouds with radii $\la 1$ pc (Lord et al. 1996; Mao et al. 2000).
If this were the case, the observed radio emission is presumably
from shock fronts in the dense gas, but some of the expansion
is in a lower density medium.
However, an LVG (large velocity gradient) analysis of the CO data
yields cloud radii of $\sim 150$ pc (Mao et al. 2000) and Shen \& Lo (1995)
identify $\sim 60$ CO ``clouds'' with half-power diameters
of $10-100$ pc.
In any case, the dense gas is likely to have complex structure.
The radio images of the remnant do show considerable structure
in the nonthermal emission (Muxlow et al. 1994).

An upper limit on the filling factor $f$ of $10^3$ cm$^{-3}$ gas
can be obtained by assuming that all of the central gas is at this
density.
Assuming a disk of diameter 600 pc, a total height of 100 pc, and
an interclump density of $n_H=10^3$ cm$^{-3}$, the gas mass is
$9.8\times 10^8 f\Msun$.
The total gas mass within a radius of 400 pc, including H$_2$, HI, and
HII, is $\sim 1.3\times 10^8 \Msun$ (G\"otz et al. 1990),
yielding $f\la 0.13$.
The filling factor appears to be low, but we expect the supernovae
to come from massive stars that are born in the molecular gas, so
that the supernovae are correlated with the presence of molecular
gas.

Although a late red supergiant phase is unlikely to significantly
affect the surrounding medium, as discussed above, photoionization
and a wind during
the main sequence phase could have an effect on the surroundings.
Draine \& Woods (1991) considered these effects and found that for
a surrounding density of $10^3$ cm$^{-3}$, stars with an initial mass
of $25\Msun$ or less have supernovae that enter the radiative blast wave
phase in that medium.
This should apply to most of the massive star supernovae in M82.

Wheeler, Mazurek,  \& Sivaramakrishnan (1980) find that a remnant 
in a medium with $n_H= 10^3$ cm$^{-3}$ becomes radiative at a radius
$R_c=0.78$ pc for a $10^{51}$ erg explosion.
For the same parameters, Draine \& Woods (1991) find $R_c=1.37$ pc;
the difference can be attributed to the fact that their cooling
function is a factor $\sim 2$ smaller than that used by Wheeler et al. (1980).
Except for the sources 41.9+57.5 and 44.0+59.6, which may be
an active galactic nucleus (Seaquist, Frayer, \& Frail 1997), 
Muxlow et al. (1994) find that
the diameters of the compact remnants in M82 are in the range 1.1 to 4 pc.
We suggest that the remnants in M82 have recently entered the radiative
phase of evolution.
The evidence in our Galaxy is that remnants which have recently
become radiative (e.g., W28, W44) have higher radio surface brightnesses for a given
diameter than those that are non-radiative  with a similar
diameter (e.g., SN 1006).

To estimate the observed X-ray fluxes, we use an analytical model for
the evolution of the supernova remnants, as in Draine \&
Woods (1991), for different values of the ambient density, $n_H$. From
the shock temperature and density we calculate the cooling rate and
luminosity of the remnant as function of time.  Using a simple
bremsstrahlung form for the spectrum and including the absorption by
the interstellar medium, we estimate the resulting X-ray luminosity in
the 0.1 -- 2 keV and 2 -- 10 keV bands. In the upper panels of Fig.
1 we show the unabsorbed luminosities in the two bands, while the
lower panels show the observed luminosities for a
  column density of $4.3\times10^{22} {\rm cm^{-2}}$ (solid lines),
and for a lower column density of $2\times10^{22} {\rm cm^{-2}}$
(dashed lines). 
Mattila \& Meikle (2001) find that the average H column density to
the remnants, including atomic and molecular H, is
$4.3\times 10^{22}$ cm$^{-2}$ ($\sigma\approx 1.7\times 10^{22}$ cm$^{-2}$).
We terminate the calculations at the point when the
internal pressure of the remnant is equal to the external pressure of
the interstellar medium, here taken to be $p/k = 10^7 {\rm~ K
~cm^{-3}}$. This mainly affects  the cases with  $n_H\la 10^2
{\rm ~cm^{-3}}$.
The peaks in the unabsorbed X-ray luminosity, especially apparent in the light
curves of the 0.1 -- 2 keV band, occur at the time of the transition
from the adiabatic to the radiative phase.

It is seen that in the 0.1 -- 2 keV band the observed luminosities are
$\sim 3\times10^{37} \ergs$ for $n_H = 10^3 {\rm cm^{-3}}$, and
considerably below this for lower densities. For the low value of the
column density the luminosity approaches $\sim 1\times10^{38}
\ergs$. The observed luminosities in the 2 -- 10 keV band are
basically unaffected by the interstellar medium absorption. For
$n_H = 10^3 {\rm cm^{-3}}$, the peak luminosity is $\sim
3\times10^{38} \ergs$, scaling roughly with the ambient density.

A {\it Chandra} image of M82  (Griffiths et al. 2000)
shows many compact X-ray sources, but their positions are 
$> 1^{\prime\prime}$ from
  the positions of the compact radio remnants, except in one  case.
Most of the sources are likely to be X-ray binaries, or supernova remnants that
are not strong radio emitters.
The luminosity limit of the image is $\sim 10^{37}\ergs$.
While especially the 2 -- 10 keV luminosities would be observable with
{\it Chandra} and in possible conflict with the small number of X-ray
sources coinciding with the compact radio sources, we note that the
peak of the X-ray flux occurs considerably before the radiative
transition when we expect an enhanced radio flux. The radio and X-ray
bright remnants may therefore represent two different populations,
with the former  more numerous because of their greater ages. 
An uncertainty in the predicted X-ray emission is that it is strongest
at a small radius when the effects of winds from the progenitor
may be a factor in the surrounding density distribution.

There are possible problems with our picture.
VLBI observations by Pedlar et al. (1999) show that the
 remnant 43.3+59.2 is expanding
at $9850\pm 1500\kms$, as opposed to the $\sim 500\kms$ that we expect.
This remnant shows a relatively constant flux (Kronberg et al. 2000),
so it belongs to the main class of remnant.
If further VLBI measurements confirm the initial result, and find
similar rapid expansion in other remnants, our model will have
to be abandoned.
Also, Muxlow et al. (1994) find the number -- diameter relation for 
the remnants to be consistent with free expansion; the number of remnants
with size less than diameter $D$ increases linearly with $D$.
  However, the relation may be due to decelerating remnants evolving in
  an interstellar medium with
density variations (Berkhuijsen 1987).
In our model, the remnants would be brightest when they first become
radiative, which is a function of density.

There are arguments, in addition to the expected rapid evolution,
against free expansion for the remnants at a velocity $\sim 10^4\kms$.
The high radio luminosities of the remnants suggests substantial
thermalization  and deceleration of the ejecta.
Huang et al. (1994) argued that the surface brightness -- diameter
relation that applies to the M82, LMC, and Galactic remnants could
be approximately understood by assuming that some constant fraction
of the supernova energy goes into relativistic particles and magnetic
fields.
The small diameter remnants are not observable in the Galaxy and LMC
because they have not interacted with a sufficiently dense medium
to thermalize the ejecta energy.

There is also the rate of formation of the radio remnants.
If the remnants expand at $10^4\kms$ and there are 28 with diameters
$<3$ pc (Huang et al. 1994), the formation rate is 1/(5 yr)
(see also Muxlow et al. 1994).
In our model, the remnants have mean expansion velocities about
an order of magnitude smaller and a correspondingly lower formation rate.
No new remnants have been found above the flux level observed
by Kronberg
et al. (1985) in the past 20 years, so the observations are more
consistent with the slow expansion model.

An individual radiative remnant is expected to have a peak total luminosity
$\sim 10^{40} \ergs$, most of which is expected to be 
at infrared wavelengths (Wheeler et al. 1980; Draine \& Woods 1991).
The problem with observing this emission is that the starburst
nucleus of M82 has an infrared luminosity $\sim 10^{44} \ergs$.
For the high absorption toward the compact remnants, the observable
emission is likely due to dust continuum emission and infrared fine
structure lines.
The [OI] 63$\mu$m line dominates the postshock cooling below 5,000 K and
should be prominent (Hollenbach \& McKee 1989).
For a newly cooling remnant in a medium with $n_H=10^3$ cm$^{-3}$
and shock velocity $v_{sh}=426\kms$ and $R=1.37$ pc (Draine \& Woods 1991),
the expected [OI] 63 $\mu$m luminosity is $1\times 10^{37}\ergs$
(Chevalier 1999).
The total observed [OI] 63 $\mu$m luminosity from the starburst
region of M82 is $\sim 1\times 10^{41}\ergs$, most of which is
thought to arise from warm neutral gas photodissociated by
radiation from OB stars (Lord et al. 1996).
The expected line emission is faint, but might be observable because
the line is broad and is expected in   localized regions.
However, we estimate that the line is somewhat below the expected detection
limit of the {\it AIRES} echelle spectrograph on {\it SOFIA}.

Greenhouse et al. (1997) have detected 6 sources of [Fe II]
1.2 and 1.6 $\mu$m line emission that they identify as shock
emission from supernova remnants.
The emission does not coincide with that from the radio remnants.
We attribute this partially to the high extinction to the radio supernova remnants.
Mattila \& Meikle (2001) find an average $A_V\approx 24$ to the
radio remnants, while Greenhouse et al. (1997) find
$A_V = 6-9$ to the [Fe II] remnants.
Taking $A_V = 7.5$ to the [Fe II] remnants and the 
extinction curve of Fitzpatrick (1999),
the additional extinction would reduce the observed flux of the
radio remnants by a factor of $13$ compared to the [Fe II] remnants.
In addition, both theoretical shock models (Hollenbach \& McKee 1989)
and observations of a Galactic remnant (Oliva et al. 1999)
indicate that the luminosity ratio  $L({\rm [Fe II] \  1.65 \mu m})/
L({\rm [O I] \ 63 \mu m}) < 1$, leading to a low expected
[Fe II]  1.65 $\mu$m line luminosity.
The same considerations indicate that the [Fe II] remnants should be strong
[O I]  63 $\mu$m line sources, easily detectable with {\it SOFIA}.
The uncertainties in this argument are that the expected shock
velocities in the radio remnants are higher than have been modeled
or observed, and the shock velocities in the [Fe II] remnants
are unknown.

Three of the [Fe II] remnants are resolved with 
diameters  of $\sim 40$ pc and the remainder are unresolved
at a diameter of 22 pc (Greenhouse et al. 1997),  an order of magnitude larger
than the radio remnants.
Expansion in a $10^3$ cm$^{-3}$ medium would lead to substantial
energy loss and a lower [Fe II] luminosity, yet the observed
high luminosity requires interaction with relatively dense gas.
We suspect that the remnants are primarily interacting with a low
density gas, and the [Fe II] emission is from radiative shock
waves in clouds.
The positions of the remnants  extend to a larger distance from
the center of M82 than do the radio remnants (Greenhouse et al. 1997).
The [Fe II] remnants should be weak nonthermal  radio sources;
based on the surface brightness -- diameter relation approximately
followed by radio remnants (Huang et al. 1994), we estimate
radio fluxes of $\sim 0.01-0.04$ mJy.
As noted by Greenhouse et al. (1997), the high radio background in M82
makes faint sources difficult to detect.

\section{OTHER GALAXIES AND SUPERWINDS}

Ulvestad \& Antonucci (1997) have studied compact radio sources in
NGC 253 with a resolution of 1 pc.
They find that both HII regions and supernova remnants are present
(based on their spectral indices), that some of the sources are
resolved, and that there is little evidence for flux variation
over a timespan of 8 yr.
The upper limits on the rate of flux decrease of the strongest
sources are $1-2$\% yr$^{-1}$.
These properties indicate that the remnants are quite similar to
those in M82.

The sources found by Smith et al. (1998) in the NW nucleus of the
Arp 220 merging system have somewhat different properties:
they are unresolved with a $1.1\times 2.9$ pc beam and they are
more luminous than the M82 sources.
Their luminosities are comparable to that of the luminous radio
supernova SN 1986J and Smith et al. (1998) identify the objects
with this class of radio source.
This requires that essentially all of the supernovae in the nuclear
region of Arp 220 produce very luminous radio supernovae.
Events like SN 1986J are thought to involve interaction with
an especially dense wind and are rare among supernova explosions.
However, the interstellar conditions in the nuclear region of Arp 220
are probably more extreme than those in M82 and may account for
the difference in the radio source properties.
The projected area covered by the radio sources is more compact:
$75\times 150$ pc in Arp 220 as compared to $100\times 600$ pc 
in M82.
The infrared luminosity in Arp 220 is $\sim 10^{12}\lsun$ as
compared to $\sim 4\times 10^{10}\lsun$ in M82, which suggests that the
supernova rate is correspondingly higher in Arp 220.
In the spherically symmetric galactic wind theory of Chevalier \& Clegg
(1985), the central pressure is $p_c\propto \dot E^{1/2} \dot M_w^{1/2}
R^{-2}$, where $\dot E$ is the total power input from supernova,
$\dot M_w$ is the mass loss rate in the galactic wind, and $R$ is
the radius of the region of power input.
The pressure in the starburst region of Arp 220 is plausibly
$\ga 10$ greater than that in M82 and the density in the interclump
region of molecular clouds may have a correspondingly higher density.
Solomon (2001) in fact finds that the mean density in the starburst
region of Arp 220 is $\sim 10^4$ cm$^{-3}$.
For $n_H=10^4$ cm$^{-3}$, the cooling radius for a supernova remnant is
$0.5$ pc when its age is $\sim 100$ yr (Draine \& Woods
1991).
We thus predict that the remnants are only somewhat smaller than
the current size limits and that they evolve more rapidly
than the remnants in M82.

In our model for the compact remnants, they are radiative and
thus do not efficiently provide mechanical energy to heat gas
in the starburst region.
In addition, our estimate of ages of the objects is larger than
in models where they expand at a high constant velocity (e.g.,
Muxlow et al. 1994) and the rate of supernova explosions is
correspondingly lower.
We hypothesize that there is a population of massive star supernovae
that are able to escape from their parent molecular cloud and
explode in a low density, hot medium ($n_H\la 1$ cm$^{-3}$).
These supernovae are responsible for driving the galactic winds
from M82 and other starburst galaxies (Chevalier \& Clegg 1985;
Heckman,  Armus,  \& Miley 1990).
Unfortunately, the remnants of these supernovae are difficult to
observe; 
Fig. 1 shows that remnants occurring in the low density component
 should have luminosities $< 10^{36} \ergs$ and
are not detectable with {\it Chandra}.
However, the initial supernovae may be directly observable
(e.g., Bregman, Temi, \& Rank 2000).
Supernovae in the low density medium may have smaller line-sight
column densities that the ones in molecular clouds.

\acknowledgments
RAC is grateful to R. O'Connell and J. Ulvestad for useful conversations,
and we thank an anonymous referee for a very helpful report.
Support for this work was provided in part by NASA grant NAG5-8232.

\clearpage

\begin{figure}[!hbtp]
\plotone{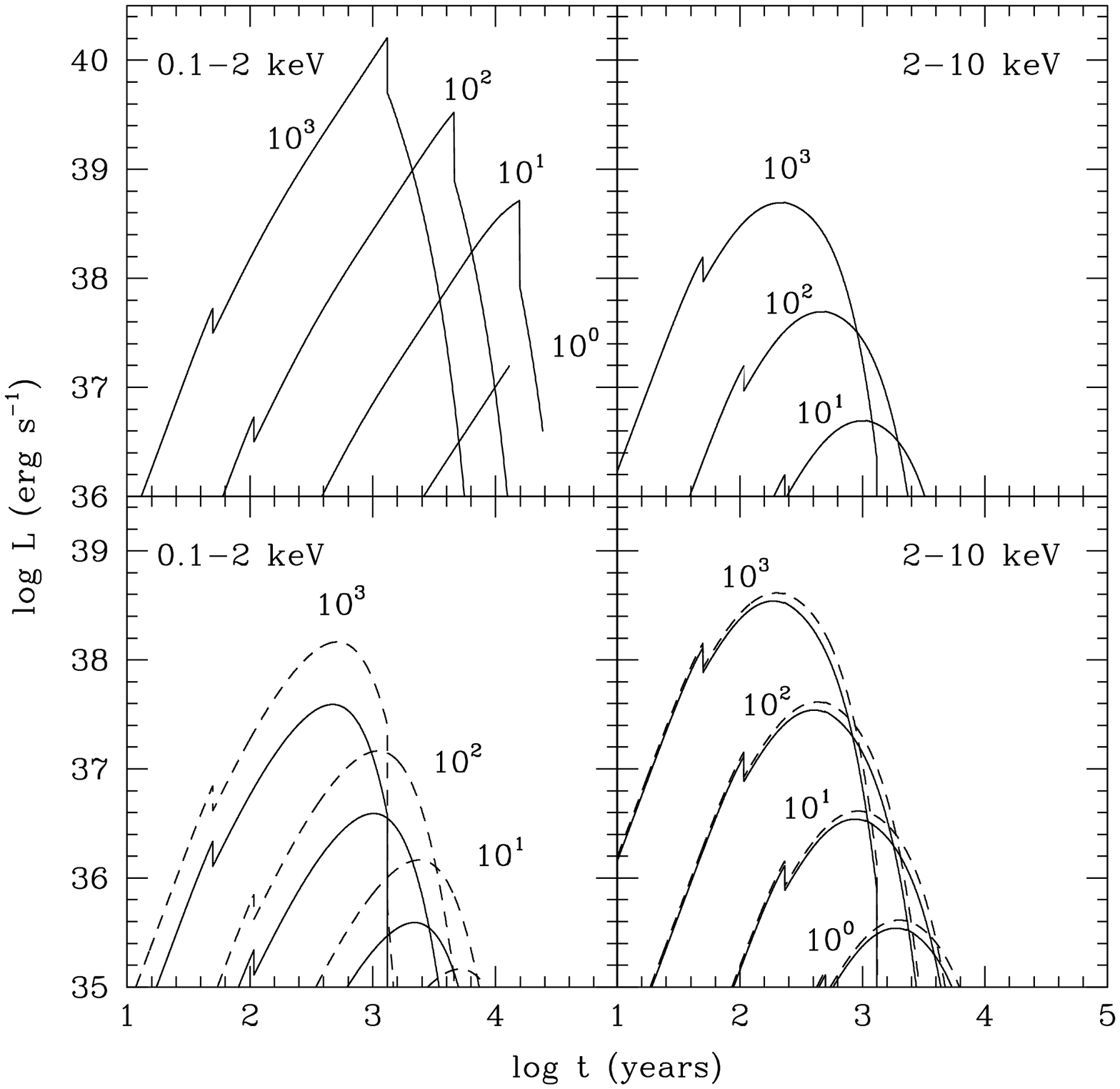}
\figcaption{X-ray luminosities in the 0.1 -- 2 keV and 2 -- 10 keV bands as a
function of time. The upper panel gives the emitted total luminosities
in the two bands for different densities. The curves for $n_0 = 1~{\rm
cm^{-3}}$ and $n_0=10 ~{\rm cm^{-3}}$ are truncated at the point where
the pressure of the interstellar medium stalls the expansion of the
remnant. The lower panel shows the observed luminosity in the two
bands corrected for interstellar absorption. The solid lines show the
results) for $N_H = 4.3\times10^{22} ~{\rm cm^{-2}}$, while the dashed
lines show the luminosities for $N_H = 2\times10^{22} ~{\rm cm^{-2}}$.}
\end{figure}

\end{document}